\documentstyle[prd,aps,preprint,tighten,epsfig]{revtex}

\begin{document}

\draft

\title{ Friedberg-Lee Symmetry for Quark Masses and Flavor Mixing }
\author{{\bf Ping Ren}
\thanks{E-mail: renp@mail.ihep.ac.cn}}
\address{ Institute of High Energy Physics, Chinese Academy of
Sciences, Beijing 100049, China }

\maketitle

\begin{abstract}
We point out a generic correlation between the Friedberg-Lee
symmetry of quark mass operators and the vanishing of quark masses.
We make phenomenological explorations on two textures of quark mass
matrices with the broken Friedberg-Lee symmetry. We present a new
pattern of quark mass matrices in agreement with current
experimental data. Both analytical and numerical results of our
calculations are discussed in detail.
\end{abstract}


\newpage

\section{Introduction}

Recent experiments \cite{PDG} have provided us with precise data on
the Cabibbo-Kobayashi-Maskawa (CKM) flavor mixing matrix \cite{CKM}
and quark masses. In order to interpret the hierarchical structure
of the observed quark mass spectrum and that of the observed flavor
mixing parameters, as well as the observed CP violation in hadronic
weak interactions, many theoretical and phenomenological models of
quark mass matrices have been proposed \cite{F and X 00,X and Z 03}.
Among them, the scenarios based on possible flavor symmetries are
particularly simple, suggestive and predictive.

In this paper, we focus our interest on a new symmetry of quark mass
operators proposed recently by Friedberg and Lee (FL) \cite{FL07}.
While for the lepton sector this symmetry has been explored a lot
\cite{FL06,FL0607}, its possible significance for the quark sector
has to be seen. Without loss of generality, the quark mass matrices
$M_{\rm u}$ (up-type) and $M_{\rm d}$ (down-type) can always be
taken to be Hermitian in the standard model or its extensions which
have no flavor-changing right-handed currents \cite{F79}. Thus the
quark mass term in the Lagrangian can be written as
\begin{eqnarray}
{\cal -L}^{}_{\rm mass} \; =~\overline{\left(u~~ c~~ t\right)}~
M^{}_{\rm u}\left( \matrix{u \cr c \cr t} \right)~+~
\overline{\left(d~~ s~~ b\right)}~M^{}_{\rm d}\left( \matrix{d\cr s
\cr b} \right) \; ,
\end{eqnarray}
where $M^{}_{\rm u}$ and $M^{}_{\rm d}$ are both Hermitian matrices.
As defined by FL \cite{FL07}, when ${\cal L}^{}_{\rm mass}$ has the
FL symmetry, it should be invariant under the transformations
\begin{eqnarray}
&&u\rightarrow u+\lambda^{1}_{\rm u}z , ~c\rightarrow
c+\lambda^{2}_{\rm u} z , ~t\rightarrow t+\lambda^{3}_{\rm u} z \; ;  \nonumber\\
&&d\rightarrow d+\lambda^{1}_{\rm d}w , s\rightarrow
s+\lambda^{2}_{\rm d} w , b\rightarrow b+\lambda^{3}_{\rm d} w \; .
\end{eqnarray}
Here $\lambda^{\rm i}_{\rm q}$ (for i = 1, 2, 3 and q = u, d) are
complex constants, while $z$ and $w$ are arbitrary space-time
independent elements of the Grassmann algebra. Considering this
arbitrariness of $z$ and $w$, we find that this invariance requires
det$M^{}_{\rm u}=0$ and det$M^{}_{\rm d}=0$, and the vector
$\lambda^{}_{\rm q}=( \lambda^{1}_{\rm q},~\lambda^{2}_{\rm
q},~\lambda^{3}_{\rm q} )^{T}$ (for q = u or d) to satisfy
$M^{}_{\rm q} \lambda^{}_{\rm q}=0$. In other words, the FL symmetry
implies  the existence of zero mass eigenvalues of $M^{}_{\rm u}$
and $M^{}_{\rm d}$. On the other hand, if there are zero mass
eigenvalues for both $M^{}_{\rm u}$ and $M^{}_{\rm d}$, then there
will be $\lambda^{}_{\rm q}=( \lambda^{1}_{\rm q},~\lambda^{2}_{\rm
q},~\lambda^{3}_{\rm q} )^{T}$ (for q = u or d) as the solution for
$~M^{}_{\rm q} \lambda^{}_{\rm q}=0~$, and ${\cal L}^{}_{\rm mass}$
will have the FL symmetry under the transformations in Eq. (2). In
one word, the FL symmetry of quark mass operators is equivalent to
the vanishing of quark mass eigenvalues. This conclusion can be
easily extended to the lepton sector.

A specific parametrization of quark mass matrices proven to be
convenient in the analysis of the FL symmetry is
\begin{eqnarray}
M^{}_{\rm q} \; = \; \left( \matrix{\gamma^{}_{\rm q}+\beta^{}_{\rm
q} |\eta^{}_{\rm q}|^2 & -\beta^{}_{\rm q}\eta^{}_{\rm
q}&-\gamma^{}_{\rm q}\zeta^{*}_{\rm q} \cr -\beta^{}_{\rm
q}\eta^{*}_{\rm q}&\beta^{}_{\rm q} + \alpha^{}_{\rm q}|\xi^{}_{\rm
q}|^2&-\alpha^{}_{\rm q}\xi^{}_{\rm q} \cr -\gamma^{}_{\rm
q}\zeta^{}_{\rm q}&-\alpha^{}_{\rm q}\xi^{*}_{\rm q}&\alpha^{}_{\rm
q}+\gamma^{}_{\rm q}|\zeta^{}_{\rm q}|^2} \right) \;
\end{eqnarray}
for q = u or d. Among the parameters, $\alpha^{}_{\rm q}$,
$\beta^{}_{\rm q}$ and $\gamma^{}_{\rm q}$ are real, while
$\xi^{}_{\rm q}$, $\eta^{}_{\rm q}$ and $\zeta^{}_{\rm q}$ are
complex
\footnote{Given real $\xi^{}_{\rm q}$, $\eta^{}_{\rm q}$ and
$\zeta^{}_{\rm q}$, Eq. (3) will reproduce the parametrization of
the $3\times 3$ real symmetric matrix in Ref. \cite{FL07}.}.
One can find
\begin{eqnarray}
{\rm det} M^{}_{\rm q} \; = \; \alpha^{}_{\rm q}\beta^{}_{\rm
q}\gamma^{}_{\rm q}~|1-\xi^{}_{\rm q}\eta^{}_{\rm q}\zeta^{}_{\rm
q}|^2  \; .
\end{eqnarray}
For ${\rm det}M^{}_{\rm q}=0$ we obtain two cases: 1)
$\alpha^{}_{\rm q}=0$, $\beta^{}_{\rm q}=0$ or $\gamma^{}_{\rm
q}=0$; and 2) $~\xi^{}_{\rm q}\eta^{}_{\rm q}\zeta^{}_{\rm q}=1$. In
the second case, if we have $\arg(\xi^{}_{\rm u})=\arg(\xi^{}_{\rm
d})$ and $\arg(\eta^{}_{\rm u})=\arg(\eta^{}_{\rm d})$, then the
matrix $U={\rm diag}~\{e^{i[\arg(\xi^{}_{\rm u})+\arg(\eta^{}_{\rm
u})]},e^{i\arg(\xi^{}_{\rm u})},1\}$ can make $M^{}_{\rm u}$ and
$M^{}_{\rm d}$ real through the unitary transformations
$U^{\dagger}M^{}_{\rm u}U$ and $U^{\dagger}M^{}_{\rm d}U$, which
implies no CP violation in the quark sector. Furthermore, if we have
$\beta^{}_{\rm u}=\gamma^{}_{\rm u}|\zeta^{}_{\rm u}|^2$ and
$\beta^{}_{\rm d}=\gamma^{}_{\rm d}|\zeta^{}_{\rm d}|^2$ together
with $\xi^{}_{\rm u}=\xi^{}_{\rm d}$ and $\eta^{}_{\rm
u}=\eta^{}_{\rm d}$, the commutator of $M^{}_{\rm u}$ and $M^{}_{\rm
d}$ will be zero, which denotes no flavor mixing in the quark sector
\footnote{One can verify this point from Eq. (9) in Sec. II, where
conditions $\xi^{}_{\rm u}=\xi^{}_{\rm d}$ and $\eta^{}_{\rm
u}=\eta^{}_{\rm d}$ lead to $\theta^{}_{\rm u}=\theta^{}_{\rm d}$
and $\phi^{}_{\rm u}=\phi^{}_{\rm d}$, and imply an identity CKM
matrix with no flavor mixing.}.
According to the experimental data about quark masses, the FL
symmetry should be broken at low energy scales. In the first case,
the texture zeros are responsible for maintaining the FL symmetry,
and breaking the texture zeros will bring nonzero masses to the
quark mass spectrum. In the second case, either $\arg(\xi^{}_{\rm
q})+\arg(\eta^{}_{\rm q})+\arg(\zeta^{}_{\rm q})\neq2n\pi$
$(n=0,\pm1,\pm2,...)$ \cite{FL07} or $|\xi^{}_{\rm q}||\eta^{}_{\rm
q}||\zeta^{}_{\rm q}|\neq1$ can help break the FL symmetry.

The main purpose of this work is to analyze two specific textures of
quark mass matrices with the broken FL symmetry, in the assumption
that $\beta^{}_{\rm u}=\gamma^{}_{\rm u}|\zeta^{}_{\rm u}|^2~$ and
$\beta^{}_{\rm d}=\gamma^{}_{\rm d}|\zeta^{}_{\rm d}|^2~$ always
hold, and all the parameters keep real before we add CP-violating
phases into the off-diagonal elements of $M^{}_{\rm u}$ and
$M^{}_{\rm d}$ to break the FL symmetry. Note that one of the two
textures has been analyzed in Ref. \cite{FL07}. The present paper is
different from the previous one not only because we point out a
generic correlation between the FL symmetry of quark mass operators
and the vanishing of quark masses but also because our discussions
are essentially new in two aspects. (1) We shall follow both
analytical and numerical procedures to explore the textures. Our
analytical results are more useful for further phenomenological
explorations than those obtained in Ref. \cite{FL07}, and our
numerical results help find the possible parameter space. (2) We
shall propose a new pattern of quark mass matrices. Although
Friedberg and Lee have demonstrated a seemingly
phenomenologically-allowed texture, we find that it is not
completely compatible with current experimental data. In contrast,
our new pattern is fully consistent with the experimental data.

The remaining parts of this paper are organized as follows. In Sec.
II, we present a phenomenological study of the quark mass matrix
texture in the FL ansatz \cite{FL07}, where CP-violating phases are
located in (2,3) and (3,2) elements of $M^{}_{\rm u}$ and $M^{}_{\rm
d}$. In Sec. III, we propose and explore a new ansatz where
CP-violating phases are located in (1,2) and (2,1) elements of
$M^{}_{\rm u}$ and $M^{}_{\rm d}$. Finally, we make a brief summary
in Sec. IV.

\vspace{0.4cm}

\section{The FL ansatz}

In our assumption, the quark mass matrix with CP invariance is (q =
u or d)
\begin{eqnarray}
M^{0}_{\rm q} \; = \; \left( \matrix{\beta^{}_{\rm q}(1+\xi^{2}_{\rm
q})\eta^{2}_{\rm q} & -\beta^{}_{\rm q}\eta^{}_{\rm q} &
-\beta^{}_{\rm q}\xi^{}_{\rm q}\eta^{}_{\rm q} \cr -\beta^{}_{\rm
q}\eta^{}_{\rm q} & \beta^{}_{\rm q} + \alpha^{}_{\rm q}\xi^{2}_{\rm
q} & -\alpha^{}_{\rm q}\xi^{}_{\rm q} \cr -\beta^{}_{\rm
q}\xi^{}_{\rm q}\eta^{}_{\rm q} & -\alpha^{}_{\rm q}\xi^{}_{\rm q} &
\alpha^{}_{\rm q}+\beta^{}_{\rm q}} \right) \; ,
\end{eqnarray}
where all the parameters are real. We introduce several new
variables by defining
\begin{eqnarray}
\xi^{}_{\rm u}=\tan \phi^{}_{\rm u},~~~~\xi^{}_{\rm d}=\tan
\phi^{}_{\rm d} ,~~~~ \eta^{}_{\rm u}=\tan \theta^{}_{\rm u} \cos
\phi^{}_{\rm u},~~~~\eta^{}_{\rm d}=\tan \theta^{}_{\rm d} \cos
\phi^{}_{\rm d}
\end{eqnarray}
with $\phi^{}_{\rm u,d}\in(-\frac{\pi}{2},\frac{\pi}{2})$ and
$\theta^{}_{\rm u,d}\in(-\frac{\pi}{2},\frac{\pi}{2})$. The
eigenvalues and eigenvectors are
\begin{eqnarray}
&&m^{0}_{ u} \; = \; 0 \; , ~~~~~~ m^{0}_{ c} \; = \; \beta^{}_{\rm
u}\sec^{2}\theta^{}_{\rm u} \; , ~~~~~~ m^{0}_{ t} \; = \;
\alpha^{}_{\rm u}\sec^{2}\phi^{}_{\rm u}+\beta^{}_{\rm u} \; ,
\nonumber \\
&&m^{0}_{ d} \; = \; 0 \; , ~~~~~~ m^{0}_{ s} \; = \; \beta^{}_{\rm
d}\sec^{2}\theta^{}_{\rm d} \; , ~~~~~~ m^{0}_{ b} \; = \;
\alpha^{}_{\rm d}\sec^{2}\phi^{}_{\rm d}+\beta^{}_{\rm d} \;
\end{eqnarray}
and
\begin{eqnarray}
|u\rangle^{}_{\rm 0} & = & \left( \matrix{ \cos \theta^{}_{\rm u}
\cr \sin\theta^{}_{\rm u} \cos \phi^{}_{\rm u} \cr \sin
\theta^{}_{\rm u} \sin \phi^{}_{\rm u}} \right) \; ,
|c\rangle^{}_{\rm 0} \; = \; \left( \matrix{-\sin \theta^{}_{\rm u}
\cr \cos \theta^{}_{\rm u} \cos \phi^{}_{\rm u} \cr \cos
\theta^{}_{\rm u} \sin \phi^{}_{\rm u}} \right) \; ,
|t\rangle^{}_{\rm 0} \; = \; \left( \matrix{0 \cr -\sin \phi^{}_{\rm
u} \cr \cos\phi^{}_{\rm u}}\right) \; , \nonumber\\
|d\rangle^{}_{\rm 0} & = & \left(\matrix{\cos \theta^{}_{\rm d}\cr
\sin\theta^{}_{\rm d} \cos \phi^{}_{\rm d} \cr \sin \theta^{}_{\rm
d} \sin \phi^{}_{\rm d}}\right) \; , |s\rangle^{}_{\rm 0} \; = \;
\left( \matrix{-\sin \theta^{}_{\rm d}\cr \cos \theta^{}_{\rm d}
\cos \phi^{}_{\rm d}\cr \cos \theta^{}_{\rm d} \sin \phi^{}_{\rm d}}
\right) \; , |b\rangle^{}_{\rm 0} \; = \; \left( \matrix{ 0 \cr
-\sin \phi^{}_{\rm d}\cr \cos \phi^{}_{\rm d}} \right) \; .
\end{eqnarray}
Here $M^{0}_{\rm u}$ and $M^{0}_{\rm d}$ can be diagonalized
respectively by unitary matrices $V^{0}_{\rm u}=\{|u\rangle^{}_{\rm
0},|c\rangle^{}_{\rm 0},|t\rangle^{}_{\rm 0}\}$ and $V^{0}_{\rm
d}=\{|d\rangle^{}_{\rm 0}, |s\rangle^{}_{\rm 0},|b\rangle^{}_{\rm
0}\}$ through the unitary transformations $V^{0\dagger}_{\rm
u}M^{0}_{\rm u}V^{0}_{\rm u}$ and $V^{0\dagger}_{\rm d}M^{0}_{\rm
d}V^{0}_{\rm d}$. Therefore without CP violation, we derive the CKM
matrix $V^{0}_{\rm CKM}$ from $V^{0}_{\rm CKM}=V^{0\dagger}_{\rm
u}V^{0}_{\rm d}$. We have
\begin{equation}
V^{0}_{\rm CKM} \; = \; \left( \matrix{ \cos \theta^{}_{\rm u} \cos
\theta^{}_{\rm d} + \sin\theta^{}_{\rm u} \sin\theta^{}_{\rm d}
\cos\phi & -\cos\theta^{}_{\rm u} \sin\theta^{}_{\rm d} +
\sin\theta^{}_{\rm u}\cos\theta^{}_{\rm d} \cos\phi &
\sin\theta^{}_{\rm u} \sin\phi \cr -\sin \theta^{}_{\rm u}\cos
\theta^{}_{\rm d} + \cos\theta^{}_{\rm u} \sin\theta^{}_{\rm d}
\cos\phi & \sin\theta^{}_{\rm u} \sin\theta^{}_{\rm d} +
\cos\theta^{}_{\rm u} \cos\theta^{}_{\rm d} \cos\phi &
\cos\theta^{}_{\rm u}\sin\phi \cr -\sin\theta^{}_{\rm d} \sin\phi &
-\cos\theta^{}_{\rm d}\sin\phi & \cos\phi} \right )\; \nonumber \\
\end{equation}
with $\phi=\phi^{}_{\rm u}-\phi^{}_{\rm d}$.

\vspace{1.0cm}

By adding complex phases into (2,3) and (3,2) elements, we obtain
$M^{}_{\rm q}$ (q = u or d) in the FL ansatz as
\begin{eqnarray}
M^{}_{\rm q} \; = \; \left( \matrix{\beta^{}_{\rm q}(1+\xi^{2}_{\rm
q})\eta^{2}_{\rm q} & -\beta^{}_{\rm q}\eta^{}_{\rm q} &
-\beta^{}_{\rm q}\xi^{}_{\rm q}\eta^{}_{\rm q} \cr -\beta^{}_{\rm
q}\eta^{}_{\rm q} & \beta^{}_{\rm q} + \alpha^{}_{\rm q}\xi^{2}_{\rm
q} & -\alpha^{}_{\rm q}\xi^{}_{\rm q}e^{ -i\chi^{}_{\rm q}} \cr
-\beta^{}_{\rm q}\xi^{}_{\rm q}\eta^{}_{\rm q} & -\alpha^{}_{\rm
q}\xi^{}_{\rm q}e^{ i\chi^{}_{\rm q}} & \alpha^{}_{\rm
q}+\beta^{}_{\rm q}} \right) \; .
\end{eqnarray}
It can be written as $M^{}_{\rm q}=M^{0}_{\rm q}+M^{'}_{\rm q}~,$
where $M^{'}_{\rm q}$ is
\begin{eqnarray}
M^{'}_{\rm q} \; = \; \alpha^{}_{\rm q}\tan\phi^{}_{\rm q}\left(~~~
\matrix{0&~~~~0&0 \cr 0&~~~~0&1-e^{ -i\chi^{}_{\rm q}} \cr
0&~~~~1-e^{ i\chi^{}_{\rm q}}&0} \right) \; .
\end{eqnarray}
The eigenvalue equations are
\begin{eqnarray}
&&m~(m-m^{0}_{ c})~(m-m^{0}_{ t}) \; = \; {\rm det}M^{}_{\rm u} \;
,\nonumber\\ &&m~(m-m^{0}_{ s})~(m-m^{0}_{ b}) \; = \; {\rm
det}M^{}_{\rm d} \;
\end{eqnarray}
with
\begin{eqnarray}
{\rm det}M^{}_{\rm q} \; = \; 2\alpha^{}_{\rm q}\beta^{\rm 2}_{\rm
q}\tan^{\rm 2}\theta^{}_{\rm q}\sin^{\rm 2}\phi^{}_{\rm
q}~(1-\cos\chi^{}_{\rm q}) \; .
\end{eqnarray}
Neglecting corrections of ${\cal O}(m^{}_{ u}/m^{}_{ c})$, ${\cal
O}(m^{}_{ u}/m^{}_{ t})$, ${\cal O}(m^{}_{ d}/m^{}_{ s})$ and ${\cal
O}(m^{}_{ d}/m^{}_{ b})$, we obtain
\begin{eqnarray}
&&\alpha^{}_{\rm u} \; \approx \; (m^{}_{ t}-m^{}_{ c}\cos^{\rm
2}\theta^{}_{\rm u})~\cos^{\rm 2}\phi^{}_{\rm u} \; , ~\beta^{}_{\rm
u} \; \approx \; m^{}_{ c}\cos^{\rm 2}\theta^{}_{\rm u} \;
,\nonumber\\
&&\alpha^{}_{\rm d} \; \approx \; (m^{}_{ b}-m^{}_{ s}\cos^{\rm
2}\theta^{}_{\rm d})~\cos^{\rm 2}\phi^{}_{\rm d} \; , ~\beta^{}_{\rm
d} \; \approx \; m^{}_{s}\cos^{\rm 2}\theta^{}_{\rm d} \;
\end{eqnarray}
from Eqs. (7), (12) and (13) in a good approximation.

In order to obtain the analytical approximations of elements of the
CKM matrix with CP violation, we take $M^{'}_{\rm q}$ as a
perturbation and introduce the perturbation matrix $G^{\rm
q}=V^{0\dagger}_{\rm q}M^{'}_{\rm q}V^{0}_{\rm q}~$. According to
Eqs. (8) and (11), $G^{\rm q}$ reads
\begin{eqnarray}
G^{\rm q}  & = &  \alpha^{}_{\rm q}\tan\phi^{}_{\rm
q}~(1-\cos\chi^{}_{\rm q}) \left( \matrix{ \sin^{\rm
2}\theta^{}_{\rm q} \sin 2\phi^{}_{\rm q} &   \sin\theta^{}_{\rm q}
\cos\theta^{}_{\rm q} \sin 2\phi^{}_{\rm q}  &  \sin\theta^{}_{\rm
q} \cos 2\phi^{}_{\rm q}   \cr   \sin\theta^{}_{\rm q}
\cos\theta^{}_{\rm q} \sin 2\phi^{}_{\rm q}  &   \cos^{\rm
2}\theta^{}_{\rm q} \sin 2\phi^{}_{\rm q}  &   \cos\theta^{}_{\rm q}
\cos 2\phi^{}_{\rm q}   \cr   \sin\theta^{}_{\rm q} \cos
2\phi^{}_{\rm q}   &    \cos\theta^{}_{\rm q} \cos 2\phi^{}_{\rm q}
&    - \sin 2\phi^{}_{\rm q} }\right) \nonumber \\ && +~~
i~\alpha^{}_{\rm q}\tan\phi^{}_{\rm q}\sin\chi^{}_{\rm q} ~\left(
\matrix{ 0 & 0 & \sin\theta^{}_{\rm q} \cr 0 & 0 &
\cos\theta^{}_{\rm q}\cr -\sin\theta^{}_{\rm q} &
-\cos\theta^{}_{\rm q} & 0 }\right)   \; .
\end{eqnarray}
Assuming small CP violation and small $\chi^{}_{\rm u,d}$, we write
$G^{\rm q}$ to the lowest order in $\chi^{}_{\rm q}$ as
\begin{eqnarray}
G^{\rm q} \; = \; i~\alpha^{}_{\rm q}\tan\phi^{}_{\rm
q}\sin\chi^{}_{\rm q} ~\left( \matrix{ 0 & 0 & \sin\theta^{}_{\rm q}
\cr 0 & 0 & \cos\theta^{}_{\rm q}\cr -\sin\theta^{}_{\rm q} &
-\cos\theta^{}_{\rm q} & 0 }\right)~~+~~{\cal O}\left( \chi^{\rm
2}_{\rm q} \right) \; .
\end{eqnarray}
The corresponding eigenvectors with the perturbation are given to
the first order in $\chi^{}_{\rm u,d}$ by
\begin{eqnarray}
|u\rangle  & = &  |u\rangle^{}_{\rm 0} ~-~ \frac{i~\alpha^{}_{\rm
u}\tan\phi^{}_{\rm u}\sin\chi^{}_{\rm u}\sin\theta^{}_{\rm
u}}{m^{}_{ u}-m^{\rm 0}_{ t}}~|t\rangle^{}_{\rm 0} \;
,\nonumber \\
|c\rangle  & = &  |c\rangle^{}_{\rm 0} ~-~ \frac{i~\alpha^{}_{\rm
u}\tan\phi^{}_{\rm u}\sin\chi^{}_{\rm u}\cos\theta^{}_{\rm
u}}{m^{}_{ c}-m^{\rm 0}_{ t}}~|t\rangle^{}_{\rm 0} \;
,\nonumber \\
|t\rangle  & = &  |t\rangle^{}_{\rm 0} ~+~ \frac{i~\alpha^{}_{\rm
u}\tan\phi^{}_{\rm u}\sin\chi^{}_{\rm u}\sin\theta^{}_{\rm
u}}{m^{}_{ t}-m^{\rm 0}_{ u}}~|u\rangle^{}_{\rm 0}
~+~\frac{i~\alpha^{}_{\rm u}\tan\phi^{}_{\rm u}\sin\chi^{}_{\rm
u}\cos\theta^{}_{\rm u}}{m^{}_{ t}-m^{\rm 0}_{ c}}
~|c\rangle^{}_{\rm 0} \;
,\nonumber \\
|d\rangle & = & |d\rangle^{}_{\rm 0} ~-~ \frac{i~\alpha^{}_{\rm
d}\tan\phi^{}_{\rm d}\sin\chi^{}_{\rm d}\sin\theta^{}_{\rm
d}}{m^{}_{ d}-m^{\rm 0}_{ b}}~|b\rangle^{}_{\rm 0} \; ,\nonumber \\
|s\rangle & = & |s\rangle^{}_{\rm 0} ~-~ \frac{i~\alpha^{}_{\rm
d}\tan\phi^{}_{\rm d}\sin\chi^{}_{\rm d}\cos\theta^{}_{\rm
d}}{m^{}_{ s}-m^{\rm
0}_{ b}}~|b\rangle^{}_{\rm 0} \; ,\nonumber \\
|b\rangle & = & |b\rangle^{}_{\rm 0} ~+~ \frac{i~\alpha^{}_{\rm
d}\tan\phi^{}_{\rm d}\sin\chi^{}_{\rm d}\sin\theta^{}_{\rm
d}}{m^{}_{ b}-m^{\rm 0}_{ d}}~|d\rangle^{}_{\rm 0}
~+~\frac{i~\alpha^{}_{\rm d}\tan\phi^{}_{\rm d}\sin\chi^{}_{\rm
d}\cos\theta^{}_{\rm d}}{m^{}_{ b}-m^{\rm 0}_{ s}}
~|s\rangle^{}_{\rm 0} \; .
\end{eqnarray}

Furthermore, noticing the hierarchical structure of the observed
quark mass spectrum as well as that of the observed flavor mixing
parameters, we arrive at the following analytical approximations
\footnote{For more details, see Appendix A.}
of the CKM matrix elements:
\begin{eqnarray}
V^{}_{ us}  & = &  -\cos\theta^{}_{\rm u} \sin\theta^{}_{\rm d} ~+~
\sin\theta^{}_{\rm u}\cos\theta^{}_{\rm d} \cos\phi \nonumber
\\ && +~ i~\sin\theta^{}_{\rm u} \cos\theta^{}_{\rm d} \sin\phi~
[~\sin\chi^{}_{\rm u} \sin\phi^{}_{\rm u} \cos\phi^{}_{\rm u} ~+~
\sin\chi^{}_{\rm d} \sin\phi^{}_{\rm d} \cos\phi^{}_{\rm d}~] \;
,\nonumber\\
V^{}_{ cb}  & = & \cos\theta^{}_{\rm u}~[~\sin\phi ~+~ i~
\cos\phi~(\sin\chi^{}_{\rm d} \sin\phi^{}_{\rm d} \cos\phi^{}_{\rm
d} - \sin\chi^{}_{\rm u} \sin\phi^{}_{\rm u} \cos\phi^{}_{\rm u})~]
\; ,\nonumber\\
V^{}_{ ub}  & = & \sin\theta^{}_{\rm u}~[~\sin\phi ~+~ i~
\cos\phi~(\sin\chi^{}_{\rm d} \sin\phi^{}_{\rm d} \cos\phi^{}_{\rm
d} - \sin\chi^{}_{\rm u} \sin\phi^{}_{\rm u} \cos\phi^{}_{\rm u})~]
\; .
\end{eqnarray}
It is well known that nine elements of the CKM matrix $V$ have six
orthogonal relations, corresponding to six triangles in the complex
plane \cite{F and X 00}. Among them, the unitarity triangle defined
by $V^{*}_{ ub}V^{}_{ ud} + V^{*}_{ cb}V^{}_{ cd} + V^{*}_{
tb}V^{}_{ td} =0$ is of particular interest for the study of CP
violation at $B$-meson factories \cite{PDG}. Three inner angles of
this triangle are commonly denoted as
\begin{eqnarray}
\alpha \;  =  \; \arg \left ( - \frac{V^{*}_{ tb}V^{}_{ td}}{V^{*}_{
ub}V^{}_{ ud}}
\right ) \;\; , \nonumber \\
\beta  \;  =  \; \arg \left ( - \frac{V^{*}_{ cb}V^{}_{ cd}}{V^{*}_{
tb}V^{}_{ td}}
\right ) \;\; , \nonumber \\
\gamma \;  =  \; \arg \left ( - \frac{V^{*}_{ ub}V^{}_{ ud}}{V^{*}_{
cb}V^{}_{ cd}} \right ) \;\; .
\end{eqnarray}
So far the angle $\beta$ has unambiguously been measured from the
CP-violating asymmetry in $B^0_d$ vs $\bar{B}^0_d\rightarrow J/\psi
K_{\rm S}$ decays \cite{Browder}, while current experimental data on
$\alpha$ and $\gamma$ are not accurate. The approximate result of
$\beta$ for the FL ansatz is
\begin{eqnarray}
\tan\beta  \;  =  \;  \tan\phi \sin\theta^{}_{\rm u}
\cos\theta^{}_{\rm d} \frac{~\sin\chi^{}_{\rm u} \sin\phi^{}_{\rm u}
\cos\phi^{}_{\rm u} ~+~ \sin\chi^{}_{\rm d} \sin\phi^{}_{\rm d}
\cos\phi^{}_{\rm d}~} {-\sin \theta^{}_{\rm u}\cos \theta^{}_{\rm d}
+ \cos\theta^{}_{\rm u} \sin\theta^{}_{\rm d} \cos\phi }  \; .
\end{eqnarray}
Given $|V^{}_{ us}|\sim0.2$, $|V^{}_{ cb}|\sim0.04$, $|V^{}_{
ub}|\sim0.004$ and $\sin2\beta\sim0.7$ \cite{PDG}, we find it
difficult for the predictions of this ansatz to agree with the
experimental data. In order to satisfy the constraint conditions
$|V^{}_{ cb}|\sim0.04$ and $|V^{}_{ ub}|\sim0.004$, $\theta^{}_{\rm
u}\sim0.1$ and $\phi\sim0.04$ must hold. Then we take into
consideration the analytical approximation of $V^{}_{ us}$, and find
that $|V^{}_{ us}|\sim0.2$ requires
\begin{center}
$|-\sin \theta^{}_{\rm u}\cos \theta^{}_{\rm d} + \cos\theta^{}_{\rm
u} \sin\theta^{}_{\rm d}\cos\phi~| ~\approx~ |-\cos\theta^{}_{\rm u}
\sin\theta^{}_{\rm d} + \sin\theta^{}_{\rm u}\cos\theta^{}_{\rm d}
\cos\phi~| ~\sim~0.2~~$.
\end{center}
However, these requirements make it impossible to obtain $\tan\beta
\sim 0.4$ in order to satisfy the constraint $\sin2\beta\sim0.7$. In
such a semi-analytical way, we find out that the texture of quark
mass matrices proposed by FL \cite{FL07} is not completely
compatible with current experimental data.

In the numerical way, we confirm that this ansatz is disfavored by
the experimental data. According to Eqs. (7), (12) and (13), by
setting six quark masses as input values, we can solve
$\alpha^{}_{\rm u,d}$, $\beta^{}_{\rm u,d}$ and $\chi^{}_{\rm u,d}$
in terms of $\theta^{}_{\rm u,d}$ and $\phi^{}_{\rm u,d}$, and put
these four variables under restraint of the four experimental
observables $|V^{}_{ us}|$, $|V^{}_{ cb}|$, $|V^{}_{ ub}|$ and
$\sin2\beta$. Our input quark masses at the $M^{}_{\rm Z}$ scale are
\cite{X. Z. and Z. 07}
\begin{eqnarray}
&& m^{}_{ u} = 1.28^{+0.50}_{-0.42}~{\rm MeV}~~, ~~m^{}_{
c}=0.624\pm0.083~{\rm GeV}~~, ~~m^{}_{ t}=172.5\pm3.0~{\rm GeV},
\nonumber \\ && m^{}_{ d}=2.91^{+1.24}_{-1.20}~{\rm MeV}~~, ~~m^{}_{
s}=55^{+16}_{-15}~{\rm MeV}~~, ~~~~~~~~~~~~m^{}_{
b}=2.89\pm0.09~{\rm GeV}. \nonumber
\end{eqnarray}
And our input experimental constraint conditions on the CKM matrix
are \cite{PDG}
\begin{center}
$~~~~~~|V^{}_{ us}|=0.2257\pm0.0021~,~$ $~~~~~|V^{}_{
cb}|=(41.6\pm0.6)\times10^{\rm -3}~,~$ $|V^{}_{
ub}|=(4.31\pm0.30)\times10^{\rm -3}~,~$ $\sin2\beta=0.687\pm0.032~$.
\end{center}
Fig. 1 illustrates the outputs of $|V^{}_{ ub}|$ and $\sin2\beta$ in
this ansatz, which have been constrained by the experimental data on
$|V^{}_{ us}|$ and $|V^{}_{ cb}|$. It is clear that the prediction
on $|V^{}_{ ub}|$ is too large to agree with the measurement
$|V^{}_{ ub}|=(4.31\pm0.30)\times10^{\rm -3}$. Hence this ansatz is
actually ruled out by current experimental data.

\vspace{0.4cm}

\section{A new ansatz}

By adding CP-violating phases into (1,2) and (2,1) elements, we
obtain our new texture of quark mass matrices as
\begin{eqnarray}
M^{}_{\rm q} \; = \; \left( ~\matrix{\beta^{}_{\rm q}(1+\xi^{2}_{\rm
q})\eta^{2}_{\rm q} & -\beta^{}_{\rm q}\eta^{}_{\rm q}e^{
-i\chi^{}_{\rm q}} & -\beta^{}_{\rm q}\xi^{}_{\rm q}\eta^{}_{\rm q}
\cr -\beta^{}_{\rm q}\eta^{}_{\rm q}e^{ i\chi^{}_{\rm q}} &
\beta^{}_{\rm q} + \alpha^{}_{\rm q}\xi^{2}_{\rm q} &
-\alpha^{}_{\rm q}\xi^{}_{\rm q} \cr -\beta^{}_{\rm q}\xi^{}_{\rm
q}\eta^{}_{\rm q} & -\alpha^{}_{\rm q}\xi^{}_{\rm q} &
\alpha^{}_{\rm q}+\beta^{}_{\rm q}}~ \right) \;
\end{eqnarray}
for q = u or d. It can be written as $M^{}_{\rm q}=M^{0}_{\rm
q}+M^{''}_{\rm q}~,$ where $M^{''}_{\rm q}~$ is
\begin{eqnarray}
M^{''}_{\rm q} \; = \; \beta^{}_{\rm q}\tan\theta^{}_{\rm
q}\cos\phi^{}_{\rm q}~\left(~~ \matrix{0&~~1-e^{ -i\chi^{}_{\rm
q}}&~~~~~~0 \cr 1-e^{ i\chi^{}_{\rm q}}&~~0&~~~~~~0 \cr
0&~~0&~~~~~~0}~~ \right) \; .
\end{eqnarray}
The equations of eigenvalues and estimations of parameters
$\alpha^{}_{\rm u,d} $ and $\beta^{}_{\rm u,d} $ are the same as
Eqs. (12)-(14) in Sec. II.

Now we introduce the perturbation matrix $H^{\rm
q}=V^{0\dagger}_{\rm q}M^{''}_{\rm q}V^{0}_{\rm q}~$ as
\begin{eqnarray}
H^{\rm q}  &=&  \beta^{}_{\rm q}\tan\theta^{}_{\rm
q}\cos\phi^{}_{\rm q}(1-e^{ -i\chi^{}_{\rm q}})~\left(
\matrix{\sin\theta^{}_{\rm q} \cos\theta^{}_{\rm q} \cos\phi^{}_{\rm
q}  &  \cos^{2}\theta^{}_{\rm q} \cos\phi^{}_{\rm q} &
-\cos\theta^{}_{\rm q} \sin\phi^{}_{\rm q} \cr
-\sin^{2}\theta^{}_{\rm q} \cos\phi^{}_{\rm q}  &
-\sin\theta^{}_{\rm q} \cos\theta^{}_{\rm q} \cos\phi^{}_{\rm q}  &
\sin\theta^{}_{\rm q} \sin\phi^{}_{\rm q} \cr 0&0&0 }\right)
\nonumber \\&& +~ \beta^{}_{\rm q}\tan\theta^{}_{\rm
q}\cos\phi^{}_{\rm q}(1-e^{ i\chi^{}_{\rm q}})~\left( \matrix{
\sin\theta^{}_{\rm q} \cos\theta^{}_{\rm q} \cos\phi^{}_{\rm q} &
-\sin^{2}\theta^{}_{\rm q} \cos\phi^{}_{\rm q} &0  \cr
\cos^{2}\theta^{}_{\rm q} \cos\phi^{}_{\rm q}  & -\sin\theta^{}_{\rm
q} \cos\theta^{}_{\rm q} \cos\phi^{}_{\rm q} &0  \cr
-\cos\theta^{}_{\rm q} \sin\phi^{}_{\rm q} & \sin\theta^{}_{\rm q}
\sin\phi^{}_{\rm q} & 0} \right) \; .
\end{eqnarray}
And the corresponding eigenvectors with the perturbation are
\begin{eqnarray}
|u\rangle & = & |u\rangle^{}_{\rm 0} ~+~ \frac{H^{\rm u}_{\rm
21}}{m^{}_{ u}-m^{\rm 0}_{ c}}~|c\rangle^{}_{\rm 0} ~+~ \frac{H^{\rm
u}_{\rm 31}}{m^{}_{ u}-m^{\rm 0}_{ t}}~|t\rangle^{}_{\rm 0}
\; ,\nonumber\\
|c\rangle & = & |c\rangle^{}_{\rm 0} ~+~ \frac{H^{\rm u}_{\rm
12}}{m^{}_{ c}-m^{\rm 0}_{ u}}~|u\rangle^{}_{\rm 0} ~+~ \frac{H^{\rm
u}_{\rm 32}}{m^{}_{ c}-m^{\rm 0}_{ t}}~|t\rangle^{}_{\rm 0}
\; ,\nonumber\\
|t\rangle & = & |t\rangle^{}_{\rm 0} ~+~ \frac{H^{\rm u}_{\rm
13}}{m^{}_{ t}-m^{\rm 0}_{ u}}~|u\rangle^{}_{\rm 0} ~+~ \frac{H^{\rm
u}_{\rm 23}}{m^{}_{ t}-m^{\rm 0}_{ c}}~|c\rangle^{}_{\rm 0}
\; ,\nonumber\\
|d\rangle & = & |d\rangle^{}_{\rm 0} ~+~ \frac{H^{\rm d}_{\rm
21}}{m^{}_{ d}-m^{\rm 0}_{ s}}~|s\rangle^{}_{\rm 0} ~+~ \frac{H^{\rm
d}_{\rm 31}}{m^{}_{ d}-m^{\rm 0}_{ b}}~|b\rangle^{}_{\rm 0}
\; ,\nonumber\\
|s\rangle & = & |s\rangle^{}_{\rm 0} ~+~ \frac{H^{\rm d}_{\rm
12}}{m^{}_{ s}-m^{\rm 0}_{ d}}~|d\rangle^{}_{\rm 0} ~+~ \frac{H^{\rm
d}_{\rm 32}}{m^{}_{ s}-m^{\rm 0}_{ b}}~|b\rangle^{}_{\rm 0}
\; ,\nonumber\\
|b\rangle & = & |b\rangle^{}_{\rm 0} ~+~ \frac{H^{\rm d}_{\rm
13}}{m^{}_{ b}-m^{\rm 0}_{ d}}~|d\rangle^{}_{\rm 0} ~+~ \frac{H^{\rm
d}_{\rm 23}}{m^{}_{ b}-m^{\rm 0}_{ s}}~|s\rangle^{}_{\rm 0} \; .
\end{eqnarray}

In view of the hierarchical structure of the observed quark mass
spectrum and that of the observed flavor mixing parameters, we
arrive at the following analytical estimations
\footnote{There are more details in Appendix B.}
:
\begin{eqnarray}
V^{}_{ us}  & = & -~\cos\theta^{}_{\rm u} \sin\theta^{}_{\rm d} ~+~
\sin\theta^{}_{\rm u} \cos\theta^{}_{\rm d} \cos\phi
\nonumber\\
&& -~\sin\theta^{}_{\rm u} \cos\theta^{}_{\rm u} \cos^{\rm
2}\phi^{}_{\rm u} ~[~\sin\theta^{}_{\rm u} \sin\theta^{}_{\rm d} +
\cos\theta^{}_{\rm u} \cos\theta^{}_{\rm d}
\cos\phi~]~[(1-\cos\chi^{}_{\rm u}) \cos2\theta^{}_{\rm u}
+i\sin\chi^{}_{\rm u}] \nonumber
\\ && +~\sin\theta^{}_{\rm d} \cos\theta^{}_{\rm d} \cos^{\rm
2}\phi^{}_{\rm d} ~[~\cos\theta^{}_{\rm u} \cos\theta^{}_{\rm d} +
\sin\theta^{}_{\rm u} \sin\theta^{}_{\rm d}
\cos\phi~]~[(1-\cos\chi^{}_{\rm d}) \cos2\theta^{}_{\rm d}
+i\sin\chi^{}_{\rm d}] \; ,\nonumber\\
V^{}_{ cb} & = & \cos\theta^{}_{\rm u}\sin\phi ~\{~1~+~ \sin^{\rm
2}\theta^{}_{\rm u}\cos^{\rm 2}\phi^{}_{\rm u}~[~(1-\cos\chi^{}_{\rm
u}) \cos2\theta^{}_{\rm u} ~+~i\sin\chi^{}_{\rm u}~]~\} \;
,\nonumber\\
V^{}_{ ub} & = & \sin\theta^{}_{\rm u}\sin\phi ~\{~1~-~ \cos^{\rm
2}\theta^{}_{\rm u}\cos^{\rm 2}\phi^{}_{\rm u}~[~(1-\cos\chi^{}_{\rm
u}) \cos 2 \theta^{}_{\rm u} ~+~i\sin\chi^{}_{\rm u}~]~\} \;
,\nonumber\\
\tan\beta  & = &  \sin\theta^{}_{\rm u} \frac{\sin\chi^{}_{\rm d}
\cos\theta^{}_{\rm d} \cos^{\rm 2}\phi^{}_{\rm d}~-~\sin\chi^{}_{\rm
u} \cos\theta^{}_{\rm u} \cos^{\rm 2}\phi^{}_{\rm u} (\cos
\theta^{}_{\rm u} \cos \theta^{}_{\rm d} + \sin\theta^{}_{\rm u}
\sin\theta^{}_{\rm d} \cos\phi) } {-\sin \theta^{}_{\rm u}\cos
\theta^{}_{\rm d} + \cos\theta^{}_{\rm u} \sin\theta^{}_{\rm d}
\cos\phi }  \; .
\end{eqnarray}
Given $|V^{}_{ us}|\sim0.2$, $|V^{}_{ cb}|\sim0.04$, $|V^{}_{
ub}|\sim0.004$ and $\sin2\beta\sim0.7$ \cite{PDG}, we find it
possible for the predictions of this new ansatz to accord with the
experimental data. For instance, if we have $\theta^{}_{\rm
u}\sim0.1$, $\theta^{}_{\rm d}\sim0.26$, $\phi^{}_{\rm u}\sim0.02$,
$\phi^{}_{\rm d}\sim-0.02$, $\chi^{}_{\rm u}=0$ and $\chi^{}_{\rm
d}\sim0.72$, the four experimental constraints can be obviously
satisfied.

Following the same procedure of numerical calculations in Sec. II,
we find out that this new texture of quark mass matrices is
consistent with current experimental data. The allowed parameter
space is given in Fig. 2, for $\phi^{}_{\rm
u,d}\in(0,\frac{\pi}{2})$ and $\theta^{}_{\rm
u,d}\in(0,\frac{\pi}{2})$. And the correlations between $m^{}_{
u,d}$ and $\chi^{}_{\rm u,d}$ are presented in Fig. 3 for
$\chi^{}_{\rm u,d}\in(0,\pi)$, which indicate that $m^{}_{ u}(m^{}_{
d})$ is physically generated from the CP-violating phase
$\chi^{}_{\rm u}(\chi^{}_{\rm d})$, as demonstrated in Eq. (13).

\vspace{0.4cm}

\section{Summary}

Keeping with the long-term interest in building phenomenological
models of fermion mass matrices based on flavor symmetries and
favored by current experimental data, we have carried out a further
study of the Friedberg-Lee (FL) symmetry of quark mass operators and
its explicit breaking. We have illustrated the generic correlation
between the vanishing masses and the FL symmetry, and classified the
ways to break the FL symmetry and obtain nonzero masses. Using
current experimental data, we have analyzed two specific patterns of
quark mass matrices: one is proposed by Friedberg and Lee, and the
other by the author. The latter is phenomenologically allowed, and
may serve as a useful example which might shed light on the
underlying dynamics responsible for the generation of fermion masses
and the origin of CP violation. While exploring other textures of
fermion mass matrices with the broken FL symmetry is still
arresting, the FL symmetry at high energy scales and the intensional
meaning of the invariance under the fermion field translations are
calling for much more attentions of theorists, indeed.

\vspace{0.5cm}

{\it Acknowledgments:} The author would like to thank Prof. Z.Z.
Xing for stimulating discussions, constant encouragement, and
reading the manuscript. He is also grateful to H. Zhang and S. Zhou
for helpful discussions. This work was supported in part by the
National Natural Science Foundation of China.

\newpage

\section*{Appendix  A}

In Sec. II, noticing Eq. (14) and neglecting corrections of ${\cal
O}(m^{}_{ u}/m^{}_{ c})$, ${\cal O}(m^{}_{ c}/m^{}_{ t})$, ${\cal
O}(m^{}_{ u}/m^{}_{ t})$, ${\cal O}(m^{}_{ d}/m^{}_{ s})$, ${\cal
O}(m^{}_{ s}/m^{}_{ b})$ and ${\cal O}(m^{}_{ d}/m^{}_{ b})$, we
obtain the eigenvectors as
\begin{eqnarray}
|u\rangle & = & |u\rangle^{}_{\rm 0} ~+~ i~\sin\chi^{}_{\rm u}
\sin\phi^{}_{\rm u} \cos\phi^{}_{\rm u}\sin\theta^{}_{\rm u}
~|t\rangle^{}_{\rm 0} \; ,\nonumber \\
|c\rangle & = & |c\rangle^{}_{\rm 0} ~+~ i~\sin\chi^{}_{\rm u}
\sin\phi^{}_{\rm u} \cos\phi^{}_{\rm u}\cos\theta^{}_{\rm u}
~|t\rangle^{}_{\rm 0} \; ,\nonumber \\
|t\rangle & = & |t\rangle^{}_{\rm 0} ~+~ i~\sin\chi^{}_{\rm u}
\sin\phi^{}_{\rm u} \cos\phi^{}_{\rm u}\sin\theta^{}_{\rm u}
~|u\rangle^{}_{\rm 0} ~+~ i~\sin\chi^{}_{\rm u} \sin\phi^{}_{\rm u}
\cos\phi^{}_{\rm u} \cos\theta^{}_{\rm u} ~|c\rangle^{}_{\rm 0} \;
,\nonumber \\
|d\rangle & = & |d\rangle^{}_{\rm 0} ~+~ i~\sin\chi^{}_{\rm d}
\sin\phi^{}_{\rm d} \cos\phi^{}_{\rm d}\sin\theta^{}_{\rm d}
~|b\rangle^{}_{\rm 0} \; ,\nonumber \\
|s\rangle & = & |s\rangle^{}_{\rm 0} ~+~ i~\sin\chi^{}_{\rm d}
\sin\phi^{}_{\rm d} \cos\phi^{}_{\rm d}\cos\theta^{}_{\rm d}
~|b\rangle^{}_{\rm 0} \; ,\nonumber \\
|b\rangle & = & |b\rangle^{}_{\rm 0} ~+~ i~\sin\chi^{}_{\rm d}
\sin\phi^{}_{\rm d} \cos\phi^{}_{\rm d}\sin\theta^{}_{\rm d}
~|d\rangle^{}_{\rm 0} ~+~ i~\sin\chi^{}_{\rm d} \sin\phi^{}_{\rm d}
\cos\phi^{}_{\rm d} \cos\theta^{}_{\rm d} ~|s\rangle^{}_{\rm 0}
.\nonumber   ~~~~~~~~~~(\rm A 1)
\end{eqnarray}

The CKM matrix is given by $V^{}_{\rm CKM}=V^{\dagger}_{\rm
u}~V^{}_{\rm d}$, where we define $V^{}_{\rm
u}=\{|u\rangle,~|c\rangle,~|t\rangle\}$ and $V^{}_{\rm
d}=\{|d\rangle, |s\rangle,~|b\rangle\}$. In this approximation,
several elements of the CKM matrix are
\begin{eqnarray}
V^{}_{ us}  & = &  V^{0}_{ us} ~+~ \sin\chi^{}_{\rm u}
\sin\phi^{}_{\rm u} \cos\phi^{}_{\rm u}\sin\theta^{}_{\rm u}
\sin\chi^{}_{\rm d} \sin\phi^{}_{\rm d} \cos\phi^{}_{\rm d}
\cos\theta^{}_{\rm d}~V^{0}_{ tb} \nonumber \\ && +~
i~[~\sin\chi^{}_{\rm d} \sin\phi^{}_{\rm d} \cos\phi^{}_{\rm d}
\cos\theta^{}_{\rm d}~V^{0}_{ ub} ~-~ \sin\chi^{}_{\rm u}
\sin\phi^{}_{\rm u} \cos\phi^{}_{\rm u} \sin\theta^{}_{\rm u}
~V^{0}_{ ts}~] \; ,\nonumber\\
V^{}_{ cd}  & = &  V^{0}_{ cd} ~+~ \sin\chi^{}_{\rm u}
\sin\phi^{}_{\rm u} \cos\phi^{}_{\rm u}\cos\theta^{}_{\rm u}
\sin\chi^{}_{\rm d} \sin\phi^{}_{\rm d} \cos\phi^{}_{\rm d}
\sin\theta^{}_{\rm d}~V^{0}_{ tb} \nonumber \\ && +~
i~[~\sin\chi^{}_{\rm d} \sin\phi^{}_{\rm d} \cos\phi^{}_{\rm d}
\sin\theta^{}_{\rm d}~V^{0}_{ cb} ~-~ \sin\chi^{}_{\rm u}
\sin\phi^{}_{\rm u} \cos\phi^{}_{\rm u} \cos\theta^{}_{\rm u}
~V^{0}_{ td}~] \; ,\nonumber\\
V^{}_{ cb}  & = &  V^{0}_{ cb} ~+~ \sin\chi^{}_{\rm u}
\sin\phi^{}_{\rm u} \cos\phi^{}_{\rm u} \cos\theta^{}_{\rm u}
\sin\chi^{}_{\rm d} \sin\phi^{}_{\rm d} \cos\phi^{}_{\rm d}
~(~\sin\theta^{}_{\rm d}~V^{0}_{ td} ~+~ \cos\theta^{}_{\rm
d}~V^{0}_{ ts}~)~ \nonumber \\ && + ~ i~[~\sin\chi^{}_{\rm d}
\sin\phi^{}_{\rm d} \cos\phi^{}_{\rm d}~(~\sin\theta^{}_{\rm
d}~V^{0}_{ cd}~+~\cos\theta^{}_{\rm d}~V^{0}_{ cs}~) ~\nonumber
\\&&-~ \sin\chi^{}_{\rm u} \sin\phi^{}_{\rm u} \cos\phi^{}_{\rm u}
\cos\theta^{}_{\rm u}~V^{0}_{ tb}~] \;
,\nonumber\\
V^{}_{ ub}  & = &  V^{0}_{ ub} ~+~ \sin\chi^{}_{\rm u}
\sin\phi^{}_{\rm u} \cos\phi^{}_{\rm u} \sin\theta^{}_{\rm u}
\sin\chi^{}_{\rm d} \sin\phi^{}_{\rm d} \cos\phi^{}_{\rm d}
~(~\sin\theta^{}_{\rm d}~V^{0}_{ td} ~+~ \cos\theta^{}_{\rm
d}~V^{0}_{ ts}~)~ \nonumber \\ && + ~ i~[~\sin\chi^{}_{\rm d}
\sin\phi^{}_{\rm d} \cos\phi^{}_{\rm d}~(~\sin\theta^{}_{\rm
d}~V^{0}_{ ud}~+~\cos\theta^{}_{\rm d}~V^{0}_{ us}~) ~\nonumber
\\&&-~ \sin\chi^{}_{\rm u} \sin\phi^{}_{\rm u} \cos\phi^{}_{\rm u}
\sin\theta^{}_{\rm u}~V^{0}_{ tb}~] \;
,\nonumber\\
V^{}_{ td}  & = &  V^{0}_{ td} ~+~ \sin\chi^{}_{\rm u}
\sin\phi^{}_{\rm u} \cos\phi^{}_{\rm u} \sin\chi^{}_{\rm d}
\sin\phi^{}_{\rm d} \cos\phi^{}_{\rm d} \sin\theta^{}_{\rm
d}~(~\sin\theta^{}_{\rm u}~V^{0}_{ ub} ~+~ \cos\theta^{}_{\rm
u}~V^{0}_{ cb}~)~ \nonumber \\ && - ~ i~[~\sin\chi^{}_{\rm u}
\sin\phi^{}_{\rm u} \cos\phi^{}_{\rm u}~(~\sin\theta^{}_{\rm
u}~V^{0}_{ ud}~+~\cos\theta^{}_{\rm u}~V^{0}_{ cd}~)~ \nonumber
\\&&-~\sin\chi^{}_{\rm d} \sin\phi^{}_{\rm d} \cos\phi^{}_{\rm d}
\sin\theta^{}_{\rm d}~V^{0}_{ tb} ~] \;
,\nonumber\\
V^{}_{ tb}  & = &  V^{0}_{ tb} ~+~ \sin\chi^{}_{\rm u}
\sin\phi^{}_{\rm u} \cos\phi^{}_{\rm u} \sin\chi^{}_{\rm d}
\sin\phi^{}_{\rm d} \cos\phi^{}_{\rm d} ~(~ \sin\theta^{}_{\rm u}
\sin\theta^{}_{\rm d}~V^{0}_{ ud} ~+~ \sin\theta^{}_{\rm u}
\cos\theta^{}_{\rm d}~V^{0}_{ us} ~ \nonumber \\ &&+~
\cos\theta^{}_{\rm u} \sin\theta^{}_{\rm d}~V^{0}_{ cd} ~+~
\cos\theta^{}_{\rm u} \cos\theta^{}_{\rm d}~V^{0}_{ cs} ~)~
\nonumber \\ && + ~ i~[~\sin\chi^{}_{\rm d} \sin\phi^{}_{\rm d}
\cos\phi^{}_{\rm d}~(~\sin\theta^{}_{\rm d}~V^{0}_{
td}~+~\cos\theta^{}_{\rm d}~V^{0}_{ ts}~) ~\nonumber
\\&&-~ \sin\chi^{}_{\rm u} \sin\phi^{}_{\rm u} \cos\phi^{}_{\rm u}~
(~\sin\theta^{}_{\rm u}~V^{0}_{ ub}~+~\cos\theta^{}_{\rm u}~V^{0}_{
cb}~)~] \; .\nonumber~~~~~~~~~~~~~~~~~~~~~~~~~~~~~~~~~~~~~~(\rm A 2)
\end{eqnarray}

Since the perturbation in terms of $\chi^{}_{\rm u}$ and
$\chi^{}_{\rm d}$ is small, we can make further estimations by
introducing the experimental data on the CKM matrix. Given current
experimental data \cite {PDG} $|V^{}_{ ud}|=0.97377\pm0.00027$,
$|V^{}_{ us}|=0.2257\pm0.0021$, $|V^{}_{
ub}|=(4.31\pm0.30)\times10^{\rm -3}$, $|V^{}_{ cd}|=0.230\pm0.011$,
$|V^{}_{ cs}|=0.957\pm0.017\pm0.093$, $|V^{}_{
cb}|=(41.6\pm0.6)\times10^{\rm -3}$, $|V^{}_{
td}|=(7.4\pm0.8)\times10^{\rm -3}$, $|V^{}_{
ts}|=(40.6\pm2.7)\times10^{\rm -3}$, and $|V^{}_{ tb}|>0.78$, we can
see the hierarchical structure of the flavor mixing parameters in
the CKM matrix and write Eq. (A2) approximately as
\begin{eqnarray}
V^{}_{ us}  & = &  -~\cos\theta^{}_{\rm u} \sin\theta^{}_{\rm d} ~+~
\sin\theta^{}_{\rm u}\cos\theta^{}_{\rm d} \cos\phi \nonumber
\\ && +~ i~\sin\theta^{}_{\rm u} \cos\theta^{}_{\rm d} \sin\phi~
[~\sin\chi^{}_{\rm u} \sin\phi^{}_{\rm u} \cos\phi^{}_{\rm u} ~+~
\sin\chi^{}_{\rm d} \sin\phi^{}_{\rm d} \cos\phi^{}_{\rm d}~] \;
,\nonumber\\
V^{}_{ cd}  & = &  -~\sin\theta^{}_{\rm u} \cos\theta^{}_{\rm d} ~+~
\cos\theta^{}_{\rm u} \sin\theta^{}_{\rm d} \cos\phi \nonumber
\\ && +~ i~\cos\theta^{}_{\rm u} \sin\theta^{}_{\rm d} \sin\phi~
[~\sin\chi^{}_{\rm u} \sin\phi^{}_{\rm u} \cos\phi^{}_{\rm u} ~+~
\sin\chi^{}_{\rm d} \sin\phi^{}_{\rm d} \cos\phi^{}_{\rm d}~] \;
,\nonumber\\
V^{}_{ cb} & = &  \cos\theta^{}_{\rm u} \sin\phi ~+~
i~\cos\theta^{}_{\rm u} \cos\phi~[~\sin\chi^{}_{\rm d}
\sin\phi^{}_{\rm d} \cos\phi^{}_{\rm d} ~-~ \sin\chi^{}_{\rm u}
\sin\phi^{}_{\rm u} \cos\phi^{}_{\rm u}~] \; ,\nonumber\\
V^{}_{ ub} & = &  \sin\theta^{}_{\rm u} \sin\phi ~+~
i~\sin\theta^{}_{\rm u} \cos\phi~[~\sin\chi^{}_{\rm d}
\sin\phi^{}_{\rm d} \cos\phi^{}_{\rm d} ~-~ \sin\chi^{}_{\rm u}
\sin\phi^{}_{\rm u} \cos\phi^{}_{\rm u}~] \; ,\nonumber\\
V^{}_{ td} & = & -\sin\theta^{}_{\rm d} \sin\phi ~+~
i~\sin\theta^{}_{\rm d} \cos\phi~[~\sin\chi^{}_{\rm d}
\sin\phi^{}_{\rm d} \cos\phi^{}_{\rm d} ~-~ \sin\chi^{}_{\rm u}
\sin\phi^{}_{\rm u} \cos\phi^{}_{\rm u}~] \; ,\nonumber\\
V^{}_{ tb} & = &  \cos\phi ~-~ i~ \sin\phi~ [~\sin\chi^{}_{\rm u}
\sin\phi^{}_{\rm u} \cos\phi^{}_{\rm u} ~+~ \sin\chi^{}_{\rm d}
\sin\phi^{}_{\rm d} \cos\phi^{}_{\rm d}~] \; .\nonumber
~~~~~~~~~~~~~~~~~~~~~~~~(\rm A 3)
\end{eqnarray}
Then we obtain
\begin{eqnarray}
-V^{*}_{ cb}V^{}_{ cd}V^{*}_{ td}V^{}_{ tb}  & = &
\cos\theta^{}_{\rm u} \sin\theta^{}_{\rm d}~|~\sin\phi + ~
i\cos\phi~[\sin\chi^{}_{\rm d} \sin\phi^{}_{\rm d} \cos\phi^{}_{\rm
d} - \sin\chi^{}_{\rm u} \sin\phi^{}_{\rm u} \cos\phi^{}_{\rm
u}]~|^{\rm ~2} \nonumber
\\ &&  \{~\cos\phi~[-\sin \theta^{}_{\rm u}\cos \theta^{}_{\rm
d} + \cos\theta^{}_{\rm u} \sin\theta^{}_{\rm d} \cos\phi] ~+
\nonumber\\ && i~\sin\theta^{}_{\rm u} \cos\theta^{}_{\rm d}
\sin\phi~ [\sin\chi^{}_{\rm u} \sin\phi^{}_{\rm u} \cos\phi^{}_{\rm
u} + \sin\chi^{}_{\rm d} \sin\phi^{}_{\rm d} \cos\phi^{}_{\rm d}]~
\}  \; .\nonumber   ~~~~~~~~~~(\rm A 4)
\end{eqnarray}
Noticing $\beta = \arg \left ( -V^{*}_{ cb}V^{}_{ cd}V^{*}_{
td}V^{}_{ tb} \right ) $, we finally obtain Eqs. (18) and (20) in
Sec. II.

\vspace{1.0cm}

\section*{Appendix  B}

In Sec. III, we notice that $M^{''}_{\rm u}~$($M^{''}_{\rm d}~$) is
proportional to $\beta^{}_{\rm u}$($\beta^{}_{\rm d}$) and of the
order of $m^{}_{ c}$($m^{}_{ s}$). Instead of assuming small CP
violation and calculating to the first order of the CP-violating
phases as in Sec. II, here we neglect the contribution of terms
${\cal O}(m^{}_{ u}/m^{}_{ c})$, ${\cal O}(m^{}_{ c}/m^{}_{ t})$,
${\cal O}(m^{}_{ u}/m^{}_{ t})$, ${\cal O}(m^{}_{ d}/m^{}_{ s})$,
${\cal O}(m^{}_{ s}/m^{}_{ b})$ and ${\cal O}(m^{}_{ d}/m^{}_{ b})$,
and obtain approximate eigenvectors
\begin{eqnarray}
|u\rangle & = & |u\rangle^{}_{\rm 0} ~-~\sin\theta^{}_{\rm u}
\cos\theta^{}_{\rm u} \cos^{\rm 2}\phi^{}_{\rm u}
~[(1-\cos\chi^{}_{\rm u}) \cos2\theta^{}_{\rm u} - i\sin\chi^{}_{\rm
u}]~|c\rangle^{}_{\rm 0} \; ,\nonumber\\
|c\rangle & = & |c\rangle^{}_{\rm 0} ~+~ \sin\theta^{}_{\rm u}
\cos\theta^{}_{\rm u} \cos^{\rm 2}\phi^{}_{\rm u}
~[(1-\cos\chi^{}_{\rm u}) \cos2\theta^{}_{\rm u} + i\sin\chi^{}_{\rm
u}]~|u\rangle^{}_{\rm 0} \; ,\nonumber\\
|t\rangle & = & |t\rangle^{}_{\rm 0} \; ,\nonumber\\
|d\rangle & = & |d\rangle^{}_{\rm 0} ~-~ \sin\theta^{}_{\rm d}
\cos\theta^{}_{\rm d} \cos^{\rm 2}\phi^{}_{\rm d}
~[(1-\cos\chi^{}_{\rm d}) \cos2\theta^{}_{\rm d} - i\sin\chi^{}_{\rm
d}]~|s\rangle^{}_{\rm 0} \; ,\nonumber\\
|s\rangle & = & |s\rangle^{}_{\rm 0} ~+~ \sin\theta^{}_{\rm d}
\cos\theta^{}_{\rm d} \cos^{\rm 2}\phi^{}_{\rm d}
~[(1-\cos\chi^{}_{\rm d}) \cos2\theta^{}_{\rm d} + i\sin\chi^{}_{\rm
d}]~|d\rangle^{}_{\rm 0}  \; ,\nonumber\\
|b\rangle & = & |b\rangle^{}_{\rm 0} \; .\nonumber
~~~~~~~~~~~~~~~~~~~~~~~~~~~~~~~~~~~~~~~~~~~~~~~~~~~~~~~~~~~~~~~~~~~~
~~~~~~~~~~~~~~~~~~~~~~~~~~~~~~(\rm B 1)
\end{eqnarray}
Several elements of the CKM matrix are given by
\begin{eqnarray}
V^{}_{ us}  & = &  V^{0}_{ us}~-~\sin\theta^{}_{\rm u}
\cos\theta^{}_{\rm u} \cos^{\rm 2}\phi^{}_{\rm u}
~[(1-\cos\chi^{}_{\rm u}) \cos2\theta^{}_{\rm u} + i\sin\chi^{}_{\rm
u}]~V^{0}_{ cs}  \nonumber
\\ && +~ \sin\theta^{}_{\rm d} \cos\theta^{}_{\rm d}
\cos^{\rm 2}\phi^{}_{\rm d} ~[(1-\cos\chi^{}_{\rm d})
\cos2\theta^{}_{\rm d} + i\sin\chi^{}_{\rm d}]~V^{0}_{ ud} ~-~
\sin\theta^{}_{\rm u} \sin\theta^{}_{\rm d} \cos\theta^{}_{\rm u}
\cos\theta^{}_{\rm d} \nonumber
\\ &&  \cos^{\rm 2}\phi^{}_{\rm u} \cos^{\rm
2}\phi^{}_{\rm d}~[(1-\cos\chi^{}_{\rm u}) \cos2\theta^{}_{\rm u} +
i\sin\chi^{}_{\rm u}]~[(1-\cos\chi^{}_{\rm d}) \cos2\theta^{}_{\rm
d} + i\sin\chi^{}_{\rm d}]~V^{0}_{ cd} \; , \nonumber\\
V^{}_{ cd} & = &  V^{0}_{ cd} ~+~ \sin\theta^{}_{\rm u}
\cos\theta^{}_{\rm u} \cos^{\rm 2}\phi^{}_{\rm u}
~[(1-\cos\chi^{}_{\rm u}) \cos2\theta^{}_{\rm u} - i\sin\chi^{}_{\rm
u}]~V^{0}_{ ud} \nonumber
\\ &&-~ \sin\theta^{}_{\rm d} \cos\theta^{}_{\rm d}
\cos^{\rm 2}\phi^{}_{\rm d} ~[(1-\cos\chi^{}_{\rm d})
\cos2\theta^{}_{\rm d} - i\sin\chi^{}_{\rm d}]~V^{0}_{ cs}~-~
\sin\theta^{}_{\rm u} \sin\theta^{}_{\rm d} \cos\theta^{}_{\rm u}
\cos\theta^{}_{\rm d} \nonumber
\\ &&  \cos^{\rm 2}\phi^{}_{\rm u} \cos^{\rm
2}\phi^{}_{\rm d}~[(1-\cos\chi^{}_{\rm u}) \cos2\theta^{}_{\rm u} -
i\sin\chi^{}_{\rm u}]~[(1-\cos\chi^{}_{\rm d}) \cos2\theta^{}_{\rm
d} - i\sin\chi^{}_{\rm d}]~V^{0}_{ us} \; , \nonumber\\
V^{}_{ cb}  & = &  V^{0}_{ cb} ~+~ \sin\theta^{}_{\rm u}
\cos\theta^{}_{\rm u} \cos^{\rm 2}\phi^{}_{\rm u}
~[(1-\cos\chi^{}_{\rm u}) \cos2\theta^{}_{\rm u} - i\sin\chi^{}_{\rm
u}]~V^{0}_{ ub}  \; ,\nonumber\\
V^{}_{ ub}  & = &  V^{0}_{ ub} ~-~\sin\theta^{}_{\rm u}
\cos\theta^{}_{\rm u} \cos^{\rm 2}\phi^{}_{\rm u}
~[(1-\cos\chi^{}_{\rm u}) \cos2\theta^{}_{\rm u} + i\sin\chi^{}_{\rm
u}]~V^{0}_{ cb}  \; ,\nonumber\\
V^{}_{ td}  & = &  V^{0}_{ td} ~-~ \sin\theta^{}_{\rm d}
\cos\theta^{}_{\rm d} \cos^{\rm 2}\phi^{}_{\rm d}
~[(1-\cos\chi^{}_{\rm d}) \cos2\theta^{}_{\rm d} - i\sin\chi^{}_{\rm
d}]~V^{0}_{ ts}  \; ,\nonumber\\
V^{}_{ tb}  & = &  V^{0}_{ tb}  \; .\nonumber
~~~~~~~~~~~~~~~~~~~~~~~~~~~~~~~~~~~~~~~~~~~~~~~~~~~~~~~~~~~~~~~~~~~~
~~~~~~~~~~~~~~~~~~~~~~~~~~~~~~~(\rm B 2)
\end{eqnarray}
Having paid attention to the hierarchical structure of the flavor
mixing parameters in the CKM matrix, we obtain further
approximations that lead to Eq. (25) in Sec. III:
\begin{eqnarray}
V^{}_{ us}  & = &  -~\cos\theta^{}_{\rm u} \sin\theta^{}_{\rm d} ~+~
\sin\theta^{}_{\rm u}\cos\theta^{}_{\rm d} \cos\phi \nonumber
\\ &&-~\sin\theta^{}_{\rm u} \cos\theta^{}_{\rm u} \cos^{\rm
2}\phi^{}_{\rm u} ~[(1-\cos\chi^{}_{\rm u}) \cos2\theta^{}_{\rm u} +
i\sin\chi^{}_{\rm u}]~[\sin\theta^{}_{\rm u} \sin\theta^{}_{\rm d} +
\cos\theta^{}_{\rm u} \cos\theta^{}_{\rm d} \cos\phi]  \nonumber
\\ && +~ \sin\theta^{}_{\rm d} \cos\theta^{}_{\rm d}
\cos^{\rm 2}\phi^{}_{\rm d} ~[(1-\cos\chi^{}_{\rm d})
\cos2\theta^{}_{\rm d} + i\sin\chi^{}_{\rm d}]~[\cos \theta^{}_{\rm
u} \cos \theta^{}_{\rm d} + \sin\theta^{}_{\rm u} \sin\theta^{}_{\rm
d} \cos\phi] \; , \nonumber\\
V^{}_{ cd}  & = &  -~\sin \theta^{}_{\rm u}\cos \theta^{}_{\rm d}
~+~ \cos\theta^{}_{\rm u} \sin\theta^{}_{\rm d} \cos\phi \nonumber
\\ && +~ \sin\theta^{}_{\rm u}
\cos\theta^{}_{\rm u} \cos^{\rm 2}\phi^{}_{\rm u}
~[(1-\cos\chi^{}_{\rm u}) \cos2\theta^{}_{\rm u} - i\sin\chi^{}_{\rm
u}]~[\cos \theta^{}_{\rm u} \cos \theta^{}_{\rm d} +
\sin\theta^{}_{\rm u} \sin\theta^{}_{\rm d} \cos\phi] \nonumber
\\ && -~ \sin\theta^{}_{\rm d} \cos\theta^{}_{\rm d}
\cos^{\rm 2}\phi^{}_{\rm d} ~[(1-\cos\chi^{}_{\rm d})
\cos2\theta^{}_{\rm d} - i\sin\chi^{}_{\rm d}]~[\sin\theta^{}_{\rm
u} \sin\theta^{}_{\rm d} + \cos\theta^{}_{\rm u} \cos\theta^{}_{\rm
d} \cos\phi] \; , \nonumber\\
V^{}_{ cb}  & = &  \cos\theta^{}_{\rm u}\sin\phi ~+~
\sin^{2}\theta^{}_{\rm u} \cos\theta^{}_{\rm u} \cos^{\rm
2}\phi^{}_{\rm u} \sin\phi ~[(1-\cos\chi^{}_{\rm u})
\cos2\theta^{}_{\rm u} - i\sin\chi^{}_{\rm u}]  \; ,\nonumber\\
V^{}_{ ub} & = & \sin\theta^{}_{\rm u} \sin\phi
~-~\sin\theta^{}_{\rm u} \cos^{2}\theta^{}_{\rm u} \cos^{\rm
2}\phi^{}_{\rm u} \sin\phi~[(1-\cos\chi^{}_{\rm u})
\cos2\theta^{}_{\rm u} + i\sin\chi^{}_{\rm u}] \; ,\nonumber\\
V^{}_{ td}  & = &  -~\sin\theta^{}_{\rm d} \sin\phi ~+~
\sin\theta^{}_{\rm d} \cos^{2}\theta^{}_{\rm d} \cos^{\rm
2}\phi^{}_{\rm d} \sin\phi~[(1-\cos\chi^{}_{\rm d})
\cos2\theta^{}_{\rm d} - i\sin\chi^{}_{\rm d}]  \; ,\nonumber\\
V^{}_{ tb} & = &  \cos\phi  \;\nonumber
~~~~~~~~~~~~~~~~~~~~~~~~~~~~~~~~~~~~~~~~~~~~~~~~~~~~~~~~~~~~~~~~~~~~
~~~~~~~~~~~~~~~~~~~~~~~~~~~~~(\rm B 3)
\end{eqnarray}
and
\begin{eqnarray}
-V^{*}_{ cb}V^{}_{ cd}V^{*}_{ td}V^{}_{ tb}  & = &
\cos\theta^{}_{\rm u} \sin\theta^{}_{\rm d} \sin^{\rm 2}\phi
\cos\phi ~[~-\sin \theta^{}_{\rm u}\cos \theta^{}_{\rm d} ~+~
\cos\theta^{}_{\rm u} \sin\theta^{}_{\rm d} \cos\phi~] \nonumber
\\ && +~ i~\sin\theta^{}_{\rm u} \cos\theta^{}_{\rm u} \sin\theta^{}_{\rm d}
\sin^{\rm 2}\phi \cos\phi ~[~\sin\chi^{}_{\rm d} \cos\theta^{}_{\rm
d} \cos^{\rm 2}\phi^{}_{\rm d}\nonumber
\\ && -~\sin\chi^{}_{\rm u}
\cos\theta^{}_{\rm u} \cos^{\rm 2}\phi^{}_{\rm u} ~(~\cos
\theta^{}_{\rm u} \cos \theta^{}_{\rm d} ~+~ \sin\theta^{}_{\rm u}
\sin\theta^{}_{\rm d} \cos\phi~)~]  \; . \nonumber ~~~~~~~~~~(\rm B
4)
\end{eqnarray}

\begin{figure}
\vspace{-1cm}
\epsfig{file=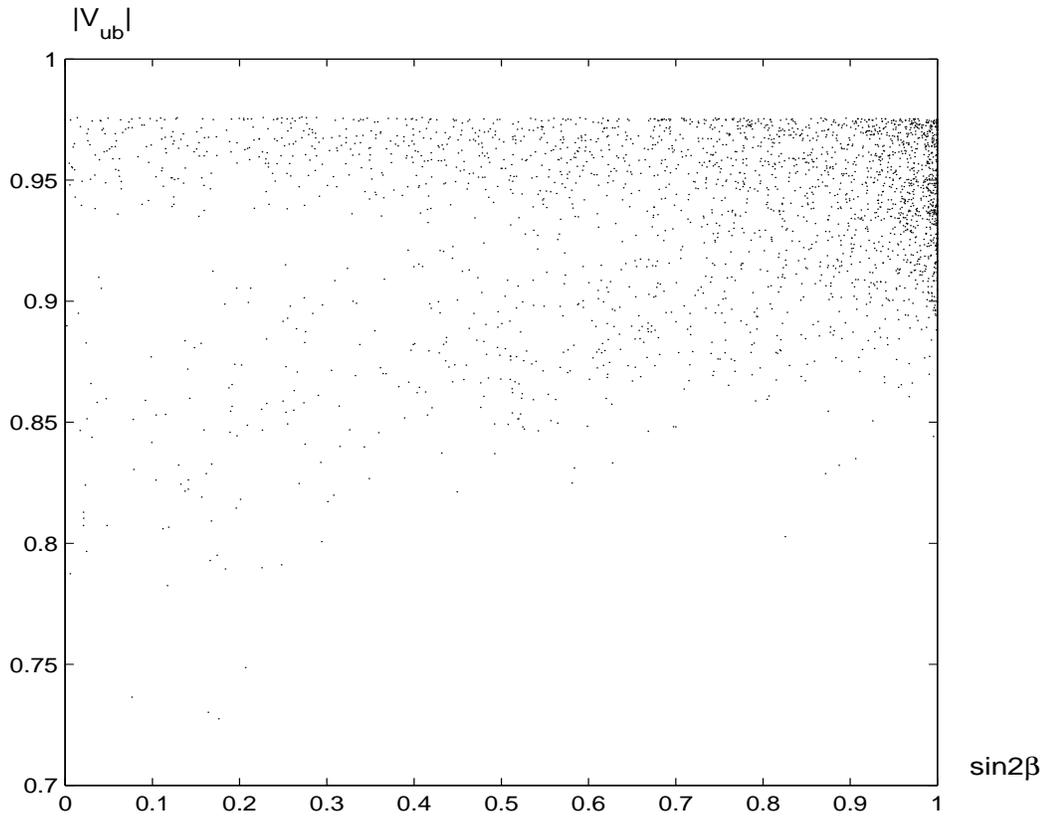, bbllx=2.5cm,bblly=12cm,bburx=17.5cm,bbury=30cm,%
width=14cm,height=14cm,angle=0,clip=90}  \vspace{4.4cm} \caption{The
ranges of the predictions on $|V^{}_{ ub}|$ vs $\sin2\beta$ when the
predictions on $|V^{}_{ us}|$ and $|V^{}_{ cb}|$ are in agreement
with current experimental data in the FL antasz in Sec. II.}
\end{figure}

\newpage

\begin{figure}
\begin{center}
\vspace{0.5cm}
\includegraphics[width=10cm,height=9cm]{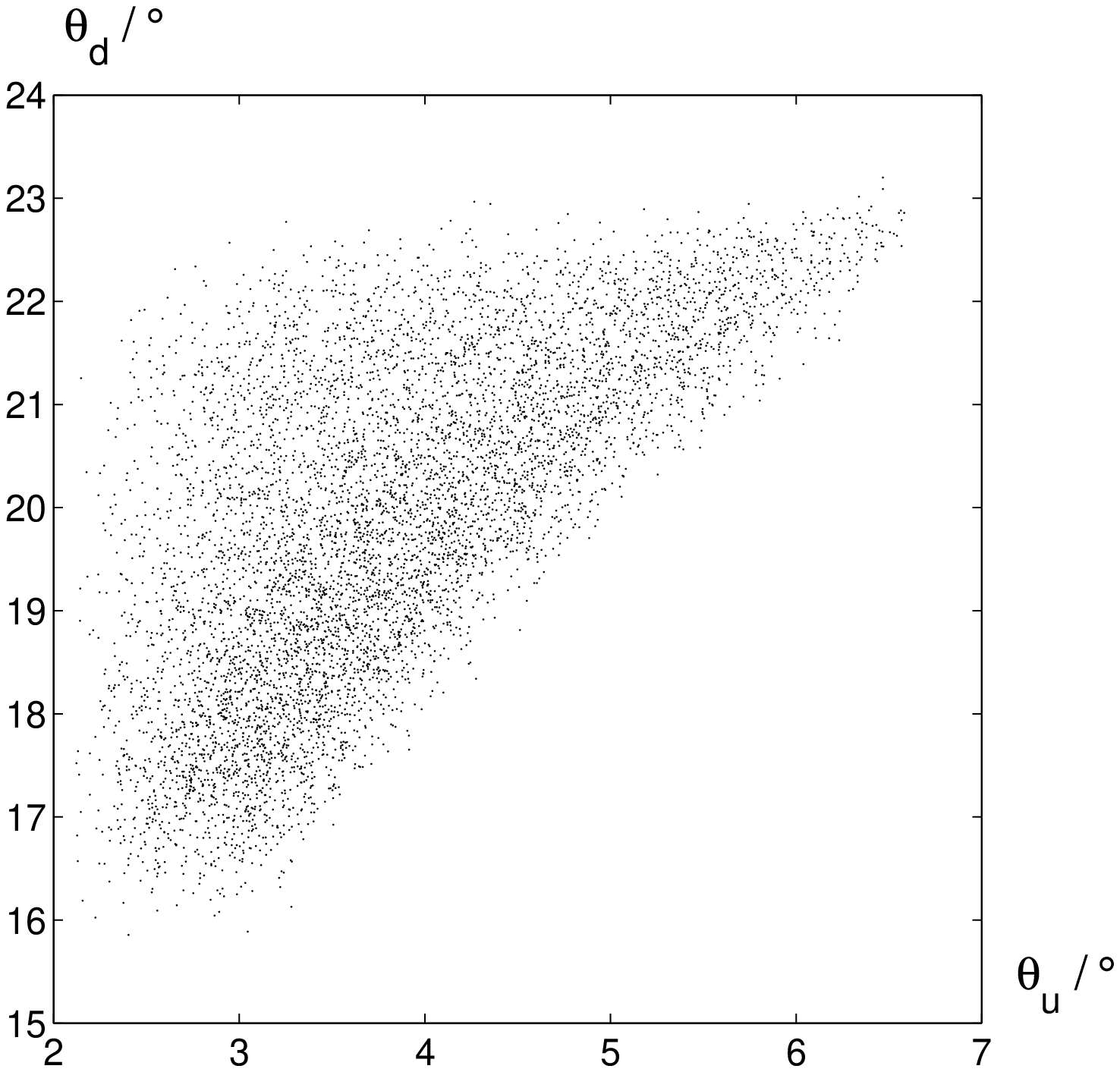}
\end{center}
\end{figure}

\begin{figure}
\begin{center}
\vspace{0.5cm}
\includegraphics[width=10cm,height=9cm]{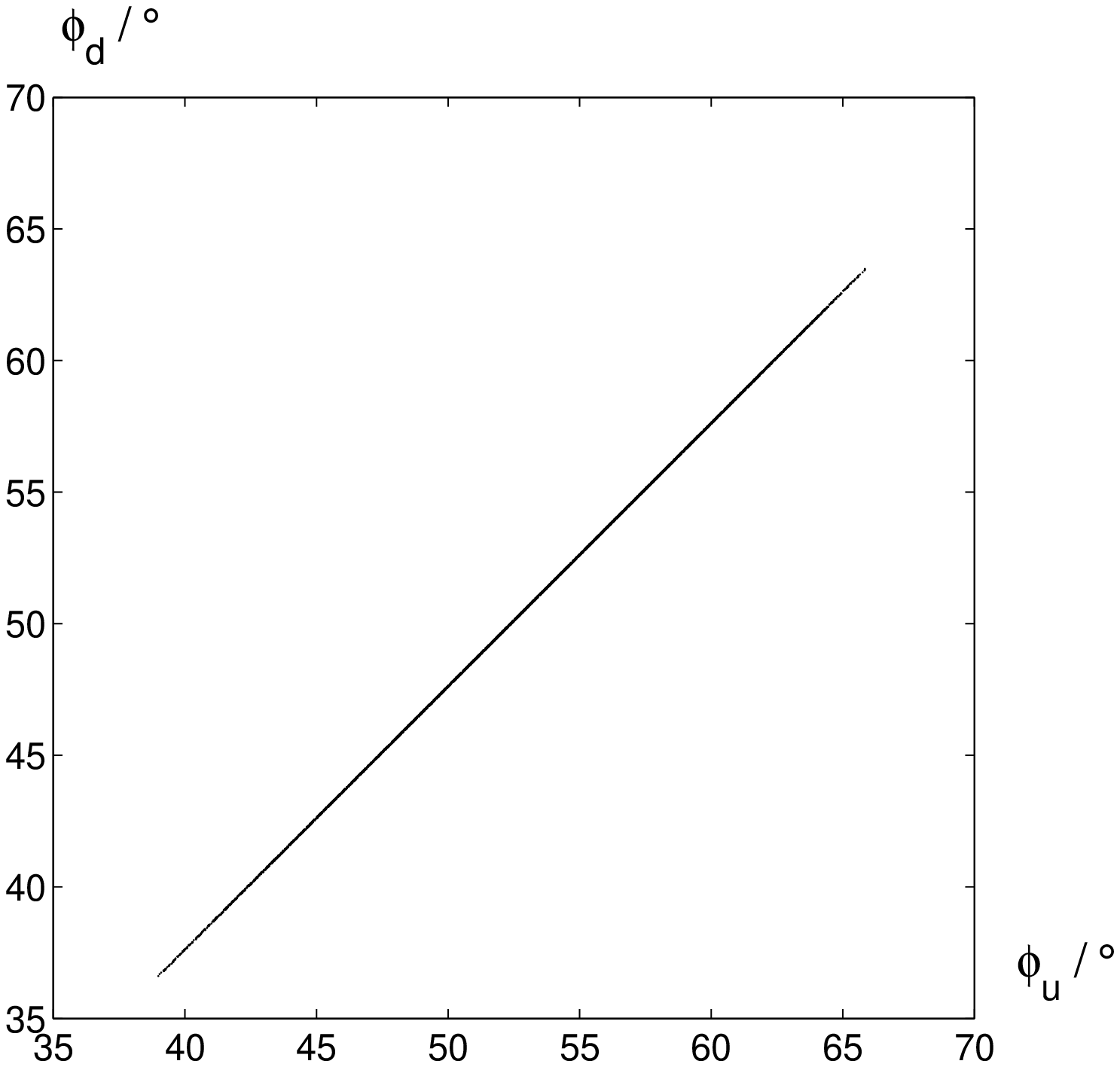}
\end{center}
\vspace{0cm} \caption{The allowed parameter space of variables
$\theta^{}_{\rm u,d}$ and $\phi^{}_{\rm u,d}$ in the new ansatz in
Sec. III.}
\end{figure}

\newpage

\begin{figure}
\begin{center}
\vspace{0.5cm}
\includegraphics[width=10cm,height=9cm]{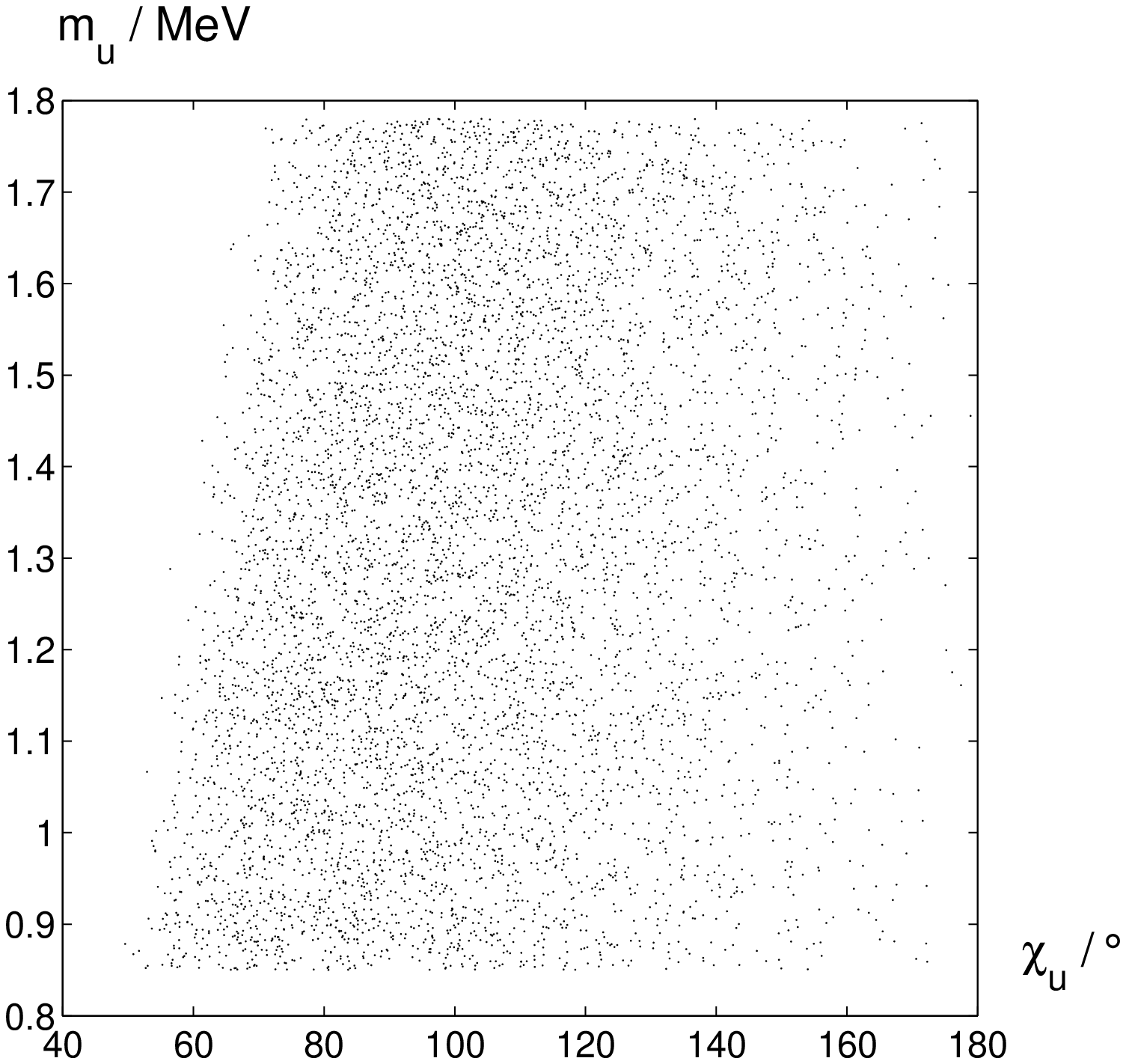}
\end{center}
\end{figure}

\begin{figure}
\begin{center}
\vspace{0.5cm}
\includegraphics[width=10cm,height=9cm]{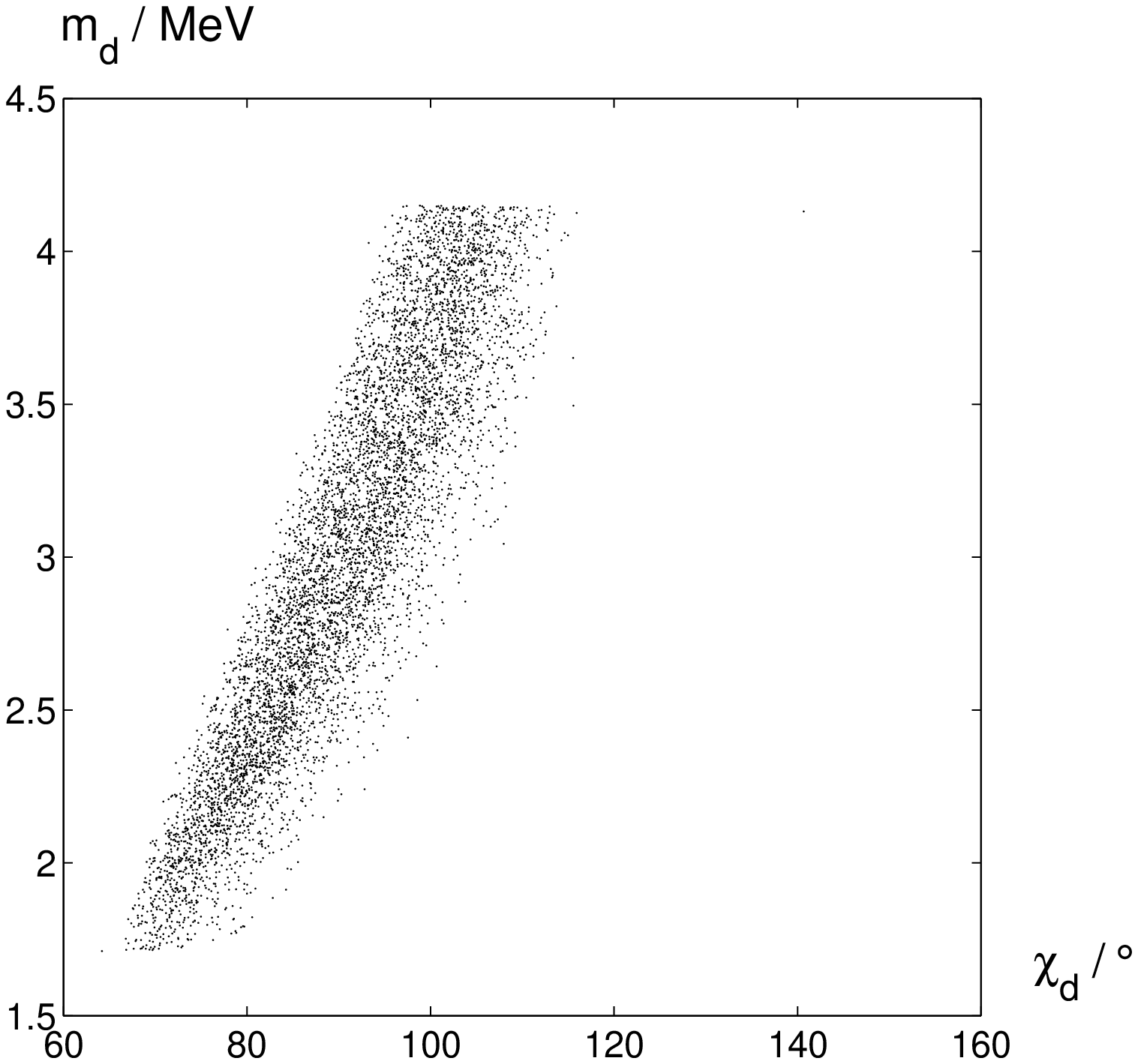}
\end{center}
\vspace{0cm} \caption{The correlations between $\chi^{}_{\rm u,d}$
and $m^{}_{ u,d}$ in the new ansatz in Sec. III.}
\end{figure}

\end{document}